# LINAC LUE200. FIRST TESTING RESULTS

S.N. Dolya, W.I. Furman, V.V. Kobets, E.M. Laziev, Yu.A. Metelkin,
V.A. Shvets, A.P. Soumbaev, JINR, Dubna, Russia
The VEPP-5 Team, BINP, Novosibirsk, Russia[1]

*Abstract*

"LUE200" – 200 MeV electron linac is being created at JINR as a driver of the pulsed neutron source "IREN" [1]. The special full-scale facilities for testing the main systems of LUE-200 (FSTF) are used at JINR, BINP, MEPhI and YerPhI [2]. The verification of the linac accelerating system is providing at the VEPP-5 preinjector constructed at BINP [3]. The accelerating system of LUE200 includes two S-band (2856 MHz) accelerating sections of 3 m long. The sections are connected with modulator based one 5045 klystron (SLAC production). There are SLED-systems for the multiplying the pulse RF power. The first results of the accelerating system test on the VEPP-5 preinjector are presented. The electron beam energy up to 92 MeV and consequently average rate of acceleration of the electron beam more than 30 MeV/m were achieved after acceleration in one section.

## 1 INTRODUCTION

In the pulsed neutron source "IREN", the well-known booster, e-γ-n scheme (electron beam + multiplying target) will be used. The target consists of tungsten converter surrounded by a plutonium ($Pu_{239}$) core.

The LUE-200 traveling wave linac conception is designed by the Budker Institute of Nuclear Physics (BINP, Novosibirsk) [1,2,3] (see Table 1).

Table 1: The parameters of LUE-200 (project)

| Beam average power | 10 – 12 kW |
|---|---|
| Electron energy | 200 MeV |
| Pulse current | 1.5 A |
| Current pulse duration | 250 ns |
| Pulse repetition rate | 150 Hz |
| Average accelerating gradient | ~ 35 MeV/m |
| Operation frequency | 2856 MHz |
| Length of the accelerating section | 3 m |
| Quantity of the accelerating sections | 2 |

The main elements of the linac and multiplying target will be mounted on three levels of the building (see Fig. 1).

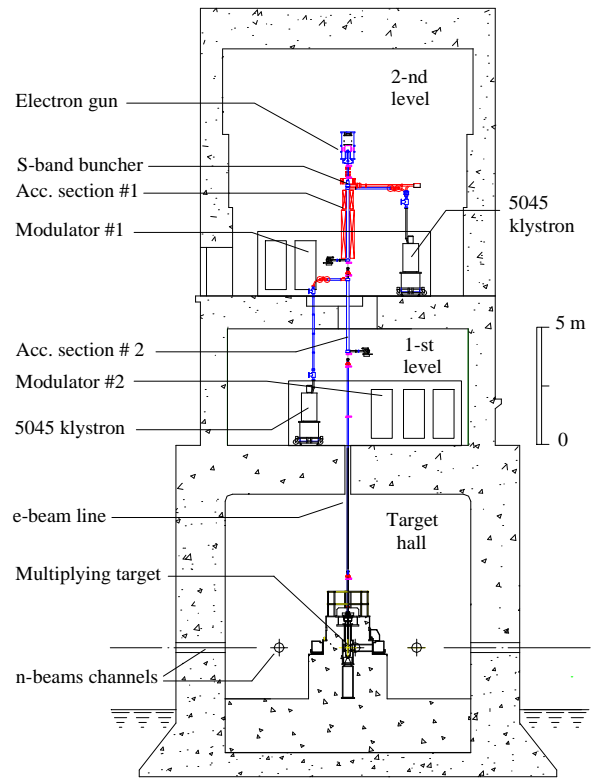

Figure 1: "IREN" neutron source layout.

## 2 EXPERIMENTAL RESULTS

An accelerating system of the linac is constructed at the Budker Institute [3,4]. The applied technology of fabrication should guarantee a stable continuous operation of the devices with a high accelerating gradient at a high RF power level. So, the aim of the test was to achieve a desired accelerating gradient ~ 35 MeV/m. The test was carried out at the initial part of the VEPP-5 preinjector [4,5]. In those experiments, the electron gun (see Table 2) and only one accelerating section of 3 m long were used. The main conditions were close to operational of the accelerator project for "IREN", namely the current of the injected beam is not less than 1.5 A for the pulse width 250 ns. This bunch duration is already the same as the one of operational part of RF power pulse.

---
[1] The full list of the VEPP-5 Team is in the references [5].

Unfortunately, the repetition rate of the setup was limited to 5 Hz because of radiation protection of environment. Another peculiarity of the test was the absence of the RF buncher (it was not ready).

The unbunched beam was steered to the input of the accelerating section. The RF pulse of up to 47 MW power is directed from 5045 klystron to the input of accelerating structure through 72 x 34 mm vacuum waveguide and SLED-compressor. The RF power collected by RF load of the accelerating section was measured both with and without beam load. The shape of RF power pulse transferred to the RF load is shown in Fig. 2. Besides, the beam current pulse profiles are given in this figure, too.

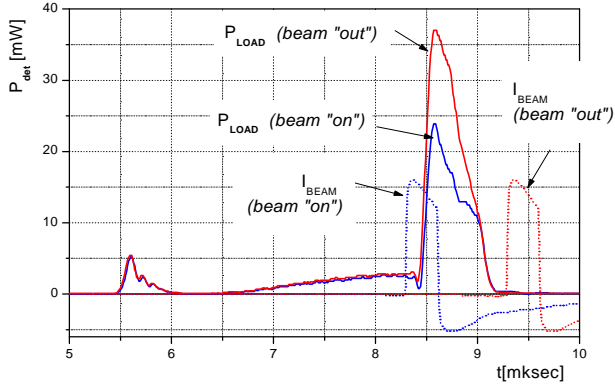

Figure 2: The shape of RF power transferred to the load both with and without beam load and the signals of current pulse from the gun wall current monitor.

The accelerating section operates with a strong current load, which leads to an essential change of the accelerating field and accelerated beam characteristics. These conditions are typical for the beam with high energy content.

Table 2: Main parameters of the electron source

| Electron gun high voltage | 120 ÷ 170 kV |
|---|---|
| Electron gun pulsed current | 1.0 ÷ 2.6 A |
| Current pulse duration | 250 ÷ 310 nsec |
| Pulse repetition rate | 1 ÷ 5 Hz |

In addition to the parameters listed above in the present series of experiments, the signals from beam multiposition collector (BMC) placed at the output of the 180°-magnetic spectrometer were also measured. The collector is performed as a set of charge collections and allows one to measure the charge with the energy resolution ≈ 1.8 MeV.

The particle energy spectrum obtained from the signals from BMC is presented in Fig. 3.

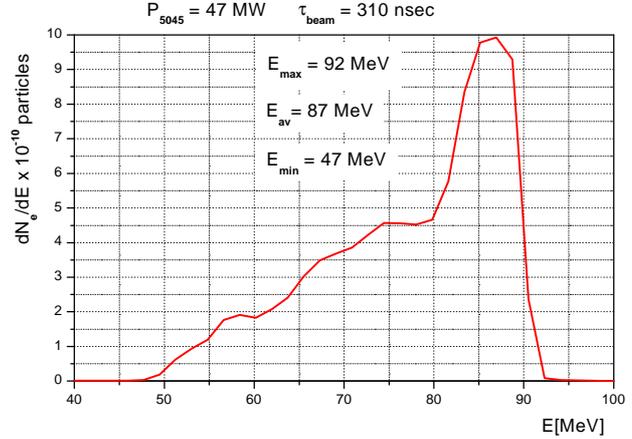

Figure 3: Particle density distribution vs. energy.

The maximum of the electron energy of the accelerated beam is 92 MeV, the minimum is 47 MeV. To our opinion, the low energy tail of the beam energy spectrum is caused by the absence of RF buncher during the test period.

The measured pulsed current from the electron gun is 2.6 A, which corresponds to the number of electrons in the pulse of $5.2 \cdot 10^{12}$ at the pulse duration of 310 nsec. The calculated number of particles after the acceleration in one section is $1.75 \cdot 10^{12}$. Thus, approximately 66% of particles were lost due to the fact the RF buncher was not available.

At the present time, after acceleration in one section without RF buncher we have following parameters:
- Klystron output RF power          47 MW.
- Total charge of the accelerated beam (in one pulse cycle)          $2.8 \cdot 10^{-7}$ C.
- Total number of particles in one pulse cycle          $1.75 \cdot 10^{12}$.
- Accelerated beam pulsed current          0.9 A.
- The average output beam energy          87 MeV.
- The beam energy content          26.1 Joule.
- The maximal output beam energy          92 MeV.
- The average acceleration rate          more than 30 MeV/m.

## 3 CONSIDERATION

Using these experimental results we can calculate the beam energy vs. the time of flight through a single accelerating section and make estimations for the situation with the main project parameters:
- Klystron 5045 output RF power          63 MW.
- Electron gun pulsed current          1.5 A.
- Current pulse duration          250 nsec.
- Repetition rate          150 Hz.

Table 3 shows the experimental and estimated parameters of this calculation. The output beam particle energy vs. time according to these parameters is shown in Fig. 4.

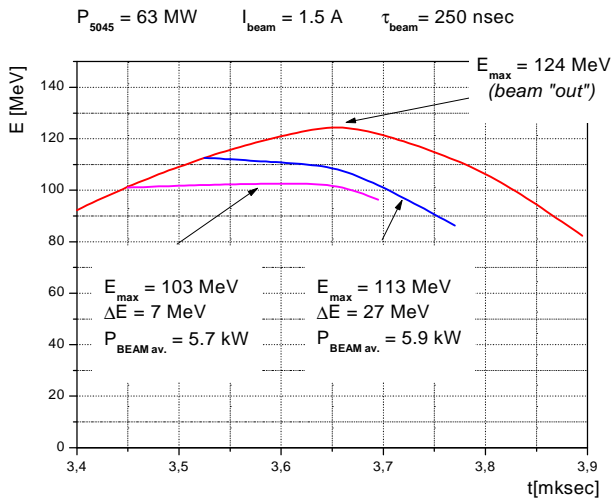

Figure 4: Output beam particle energy vs. time.

Table 3: The experimental and estimated parameters according to the beam acceleration in a single section

|  | Experiment | Estimation |
|---|---|---|
| Output 5045 klystron RF power | 47 MW | 63 MW |
| Repetition rate | 2.5 Hz | 150 Hz |
| Beam pulse current | 0.9 A | 1.5 A |
| Bean pulse duration | 310 nsec | 250 nsec |
| Average electron beam energy | 87 MeV | 103 MeV |
| Beam energy content | 26.1 Joule | 38.6 Joule |
| Average beam power | - | 5.8 kW |

## 4 CONCLUSION

An average accelerating gradient of more than 30 MeV/m was achieved in the conditions close to the "IREN" project. The experimental data allow us to expect with confidence that the high project power of the electron beam in the designed LUE-200 linac will be obtained.